\documentstyle[12pt]{article}
\def\al{\alpha}
\def\be{\beta}
\def\ga{\gamma}
\def\de{\delta}
\def\ep{\epsilon}

\def\th{\theta}

\def\si{\sigma}

\def\ph{\phi}

\def\ch{\chi}
\def\ps{\psi}

\def\Ga{\Gamma}
\def\De{\Delta}

\def\Om{\Omega}

\def\fr#1#2{{{#1} \over {#2}}}

\def\vev#1{\langle {#1}\rangle}
\def\bra#1{\langle{#1}|}
\def\ket#1{|{#1}\rangle}

\def\half{{\textstyle{1\over 2}}}
\def\frac#1#2{{\textstyle{{#1}\over {#2}}}}

\def\lsim{\mathrel{\rlap{\lower4pt\hbox{\hskip1pt$\sim$}}
    \raise1pt\hbox{$<$}}}
\def\gsim{\mathrel{\rlap{\lower4pt\hbox{\hskip1pt$\sim$}}
    \raise1pt\hbox{$>$}}}
\def\sqr#1#2{{\vcenter{\vbox{\hrule height.#2pt
         \hbox{\vrule width.#2pt height#1pt \kern#1pt
         \vrule width.#2pt}
         \hrule height.#2pt}}}}

\newcommand{\beq}{\begin{equation}}
\newcommand{\eeq}{\end{equation}}
\newcommand{\bea}{\begin{eqnarray}}
\newcommand{\eea}{\end{eqnarray}}
\newcommand{\rf}[1]{(\ref{#1})}

\renewenvironment{thebibliography}[1]
 { \rm
   \begin{list}{\arabic{enumi}.}
    {\usecounter{enumi} \setlength{\parsep}{0pt}
     \setlength{\itemsep}{3pt} \settowidth{\labelwidth}{#1.}
     \sloppy
    }}{\end{list}}

\begin{document}
\titlepage

\begin{flushright}
{COLBY-95-07\\}
{IUHET 311\\}
{August 1995\\}
\end{flushright}
\vglue 1cm

\begin{center}
{{\bf KEPLERIAN SQUEEZED STATES AND RYDBERG WAVE PACKETS \\}
\vglue 1cm
{Robert Bluhm,$^a$ V. Alan Kosteleck\'y,$^b$ and
Bogdan Tudose$^b$\\}
\bigskip
{\it $^a$Physics Department\\}
\medskip
{\it Colby College\\}
\medskip
{\it Waterville, ME 04901, U.S.A.\\}
\bigskip
{\it $^b$Physics Department\\}
\medskip
{\it Indiana University\\}
\medskip
{\it Bloomington, IN 47405, U.S.A.\\}

}
\vglue 0.8cm

\end{center}

{\rightskip=3pc\leftskip=3pc\noindent
We construct minimum-uncertainty solutions
of the three-dimensional Schr\"odinger equation
with a Coulomb potential.
These wave packets are localized in radial
and angular coordinates
and are squeezed states in three dimensions.
They move on elliptical keplerian trajectories
and are appropriate for the description of
the corresponding Rydberg wave packets,
the production of which
is the focus of current experimental effort.
We extend our analysis to incorporate
the effects of quantum defects in
alkali-metal atoms,
which are used in experiments.

}

\vskip 1truein
\centerline{\it Accepted for publication in Physical Review A}

\vfill
\newpage

\baselineskip=20pt

{\bf\noindent I. INTRODUCTION}
\vglue 0.4cm

At an early stage in the development of quantum mechanics,
Schr\"odinger attempted to find nonspreading
wave-packet solutions for a quantum-mechanical
particle evolving along a classical trajectory.
He succeeded for the case of a particle subject to a
harmonic-oscillator potential,
obtaining a solution
that follows the classical oscillatory motion without changing shape
\cite{schr}.
This solution is now called a coherent state
\cite{example}.
Schr\"odinger also attempted to find analogous
solutions for the Coulomb potential,
without success
\cite{wavemech}.
Many authors since have discussed this issue,
and it is now known that there are no exact
coherent states for the Coulomb problem
\cite{brown,mostowski,nieto,mcab,gay,nau1}.

Although the original
Schr\"odinger problem for the Coulomb potential
has no solution,
one can nonetheless obtain minimum-uncertainty
wave packets exhibiting many features of
the corresponding classical motion.
For example,
it has recently been shown that a type of squeezed state
called an elliptical squeezed state (ESS)
is a minimum-uncertainty wave packet
for the planar Coulomb problem
\cite{2dss}.
An ESS is localized in both the radial and
angular coordinates,
and it travels along a keplerian ellipse in two dimensions.
The radial part of the initial solution is a radial
squeezed state (RSS),
which minimizes the uncertainty relation for a set of
radial variables that express the Coulomb problem in a form
similar to that of an oscillator
\cite{rss,rssqdt}.
The angular part of the initial solution is
a circular squeezed state (CSS),
which minimizes the uncertainty relation between
the angular momentum $L$ and
a suitable angular-coordinate operator
\cite{2dss,kt}.

The primary goal of this paper is to extend the ESS
construction to three dimensions and thereby to
provide a class of wave packets
coming as close as possible to
a solution of the original Schr\"odinger problem
for the Coulomb potential.
We call these wave packets
\it keplerian squeezed states (KSS). \rm
We show that the KSS are
minimum-uncertainty wave packets
that are localized in all three dimensions
and that travel along a keplerian ellipse.
Three-dimensional wave packets of this kind
are of particular interest at present
because experiments using short-pulsed lasers are
attempting to produce Rydberg wave packets that move
along elliptical orbits.

To date,
experiments that have detected the classical
oscillation of a wave packet in a Coulomb potential
have been performed using purely radial wave packets
consisting of a superposition of $n$ states with $l$ fixed,
typically, to a p state
\cite{tenWolde,yeazell1,yeazell2,yeazell3,meacher,wals}.
These states follow the initial classical motion,
but also exhibit distinctive quantum-mechanical features.
For example,
the uncertainty product oscillates periodically
as a function of time.
This is a characteristic of squeezed states
\cite{kim},
and indeed the RSS provide a good analytical description of wave
packets of this type.
The wave packets also exhibit quantum interference effects,
as they undergo a series of collapses
and full/fractional revivals and superrevivals
\cite{ps,az,ap,nau2,detunings,sr}.

To generate a three-dimensional wave packet
localized in radial and both angular coordinates,
a superposition of $n$, $l$, and $m$ levels must be created.
This requires the presence of
additional fields during the excitation process.
One proposal for achieving this
involves using a short electric pulse to
convert an angular state into a localized Rydberg wave
packet moving on a circular orbit
\cite{gns}.
An additional weak electric field could then distort the
orbit into an ellipse of arbitrary eccentricity.
We expect the motion of these wave packets to be
well described by the KSS we present here.

The construction of the KSS
requires the identification of angular operators
appropriate to a problem with spherical symmetry.
These can then be used to obtain a class of squeezed states,
called spherical squeezed states (SSS),
that minimize uncertainty products for the angular variables.
This construction and the basic properties of the SSS
are presented in Sec.\ II.

Section III gives the construction of the KSS.
At the initial time,
the KSS may be written as products of
RSS with SSS.
The resulting wave packet is localized in
all three dimensions and consists of
a superposition over the three quantum numbers $n$, $l$, and $m$.
In this section,
we also calculate expectation values for physical quantities
characterizing the initial wave packet,
and we discuss its time evolution.

Section IV outlines how the KSS can be
generalized to include the effects of quantum defects.
This is necessary because experiments performed
on Rydberg wave packets generally use
alkali-metal atoms rather than hydrogen.

Section V contains a summary of our results.
Some mathematical formulae used in deriving expectation values
in Sec.\ II are given in the Appendix.
Throughout this paper,
we use atomic units with
$\hbar = m = e = 1$.

\vglue 0.6cm
{\bf\noindent II. SPHERICAL SQUEEZED STATES}
\vglue 0.4cm

In this section,
we obtain the spherical squeezed states.
These are squeezed states for angular coordinates
on a unit sphere.
Section IIA discusses the choice of angular variables
used in the construction.
Section IIB gives the derivation of the SSS,
and Sec.\ IIC examines their properties.

\vglue 0.6cm
{\bf\noindent A. Quantum Angular Variables on the Sphere}
\vglue 0.4cm

Classically,
the standard angular coordinates
on the unit sphere are
the spherical polar coordinates $\th$ and $\ph$,
which can be taken as defined
in the intervals $0 \le \th \le \pi$ and
$-\pi \le \ph < \pi$,
respectively.
At the quantum level,
however,
$\th$ are $\ph$ are not
well suited as coordinate operators.
Some of the difficulties
stem from problems already appearing in the planar case.
For example,
if $\ph$ is taken as periodic,
the discontinuity at $\ph = \pi$
causes complications when $\ph$ is acted upon by
derivative operators such as the $z$ component $L_3$
of the angular momentum $\vec L$.
The alternative assumption of a continuous angle $\ph$
in the interval $-\infty < \ph < \infty$
also introduces complications because
this variable is not periodic and hence is not observable
in the Hilbert space for which $L_3 = -i \partial_\ph$ is hermitian.
In the present instance,
the use of $\ph$ in conjunction with $\th$ on the unit sphere
raises further complexities because $\ph$ is not well defined
at the poles $\th = 0, \pi$.

Difficulties with angular coordinates
on the circle have been widely discussed in the literature
\cite{cn}.
In addressing the Schr\"odinger problem
for the planar Coulomb potential
\cite{2dss}
and in a related work on minimum-uncertainty angular packets
\cite{kt},
we followed ref.\ \cite{wl}
in circumventing these and other difficulties
by replacing $\ph$
with \it two \rm quantum coordinate operators,
$\sin\ph$ and $\cos\ph$,
which are both continuous and periodic.
Use of these variables permitted a completely satisfactory
resolution to the essential issues.

In the present work,
we follow a similar strategy.
We introduce three
angular-coordinate operators
$a_j$, $j = 1,2,3$,
via their matrix elements
obtained by insertion of the three
continuous and periodic coordinate functions
$f_j = (\sin\th \cos\ph$, $\sin\th \sin\ph$, $\cos\th)$,
respectively,
in the Hilbert-space inner product:
\beq
\bra{\ps_1}a_j \ket{\ps_2}=
\int_{0}^{2 \pi} \int_{0}^{\pi} d\Om
{}~\ps_1^* f_j\ps_2
\quad ,
\label{inner}
\eeq
where $d\Om$ is an element of solid angle on the sphere.
They obey the relation
$\sum_j a_j^2 = 1$.
Classically,
they are a natural choice corresponding to the identification
$(x,y,z) \rightarrow (\sin\th \cos\ph, \sin\th \sin\ph, \cos\th)$,
which uniquely covers the unit sphere.
Moreover,
this choice reduces
in the limit $\th \rightarrow \pi/2$
to the planar case.

The quantum operators $a_j$
and the components $L_k$
of the angular-momentum operator $\vec L$
obey the commutation relations
\beq
[a_j,L_k] = i \ep_{jkl} a_l
\quad .
\label{cr}
\eeq
Each of these commutators has an
associated uncertainty relation
that must be considered in the search
for minimum-uncertainty solutions.

The number of nontrivial relations
can be substantially reduced
by stipulating certain initial conditions
on the desired wave packet.
Without loss of generality,
we take the initial minimum-uncertainty
configuration to be a wave packet localized
about the positive $x$ axis.
Since we seek a packet moving along an orbit in the
$x$-$y$ plane,
we also assume the initial shape
is reflection-symmetric about this plane.
It is also physically reasonable to suppose
that the initial packet is reflection symmetric
about the $x$-$z$ plane.
These requirements impose the coordinate conditions
\beq
\vev{a_2} = \vev{a_3} = 0
\quad , \qquad
\vev{a_1} > 0
\quad
\label{cond}
\eeq
and the angular-momentum conditions
\beq
\vev{L_1} = \vev{L_2} = 0
\quad .
\label{cond4}
\eeq

The conditions \rf{cond}
leave only two nontrivial uncertainty relations
to be considered.
They are
\beq
\De a_2 \, \De L_3 \ge \fr 1 2 \vert
\vev{a_1} \vert
\quad
\label{ur1}
\eeq
and
\beq
\De a_3\, \De L_2 \ge \fr 1 2 \vert
\vev{a_1} \vert
\quad .
\label{ur2}
\eeq

\vfill\eject

{\bf\noindent B. Construction of SSS}
\vglue 0.4cm

The CSS
are defined as states minimizing
the relation \rf{ur1} in the $x$-$y$ plane
\cite{2dss}.
Requiring that the SSS reduce
to the CSS in the planar limit $\th \rightarrow \pi/2$
means that \rf{ur1}
must be minimized in three dimensions too.
Our construction of the SSS in this section
therefore begins with this minimization,
after which we return to the remaining relation
\rf{ur2}.

A wave function $\ch (\th,\ph)$ minimizing
Eq.\ \rf{ur1}
obeys the differential equation
\beq
a_2 \, \ch(\th,\ph) = -i \fr 1 \de
\left( L_3 - \vev{L_3} \right) \ch(\th,\ph)
\quad ,
\label{de}
\eeq
where the squeezing $\de$ in the angular coordinates is
a real constant given by
\beq
\fr 1 \de \equiv \fr {\De a_2} {\De L_3}
= \fr {2(\De a_2)^2} {\vert \vev{a_1}
\vert} \ge 0
\quad .
\label{squeeze}
\eeq
As expected,
setting $\th \to \pi/2$ reduces
Eqs.\ \rf{de} and \rf{squeeze}
to the defining equations for the CSS.
Fixing $\vev{L_3} = \be$,
a real parameter,
Eq.\ \rf{de}
becomes
\beq
\left( \partial_\ph + \de \sin\th \sin\ph - i \be \right)
\ch(\th,\ph) = 0
\quad .
\label{de2}
\eeq

The solution to this equation is
\beq
\ch(\th,\ph) = N \exp \left( \de \sin\th \cos\ph
+ i \be \ph + f(\th) \right)
\quad ,
\label{gsss}
\eeq
where $f(\th)$ is an undetermined complex function of
$\sin\th$ and $\cos\th$.
For $\ch(\th,\ph)$ to be single valued,
$\be$ must be an integer.
This follows from the general result
\cite{kt}
that minimum-uncertainty
angular wave packets must have integer expectation values
$\vev{L_3}$ of the angular momentum.

We next address the issue of the role
of the second uncertainty relation
\rf{ur2}.
An immediate question is whether the function
$f(\th)$ may be chosen so
that $\ch(\th,\ph)$ also minimizes
\rf{ur2}.
This would require that,
in addition to \rf{de2},
$\ch(\th,\ph)$ must satisfy the differential equation
\beq
\cos\th \, \ch(\th,\ph) = i \fr 1 {\de^\prime}
\left( -i \cos\ph \partial_\th
+ i \cot\th \sin\ph \, \partial_\ph \right)
\ch(\th,\ph)
\quad ,
\label{ur2eq}
\eeq
where $\de^\prime$ is a second squeezing parameter defined by
$1/\de^\prime = \De a_3/ \De L_2$.
However,
substituting \rf{gsss} into \rf{ur2eq}
produces an equality
only if $\de = \de^\prime$,
$\be = 0$, and $f(\th)$ is a constant.
Physically,
this means the squeezing ratios are equal,
$\De a_2/ \De L_3 = \De a_3/ \De L_2$,
and that the mean angular momentum vanishes,
$\vev{L_1} = \vev{L_2} = \vev{L_3} = 0$.
However,
this is unacceptable
because we seek solutions with a nonzero value for $\vev{L_3}$.
Indeed,
the two uncertainty relations
\rf{ur1} and \rf{ur2}
can be satisfied simultaneously only for states
symmetric under rotations about the $x$ axis.
The rotational symmetry must therefore be broken,
and the relation \rf{ur2} cannot be minimized.

One simple way to break the rotational symmetry would be
to keep $\vev{L_3} \ne 0$
in Eq.\ \rf{gsss} but to set $f(\th) \equiv 0$.
The function $\ch(\th,\ph)$ then would take a simplified form
that reduces directly to a CSS as $\th \rightarrow \pi/2$.
This choice has the added elegant feature
that expectation values would take the same form
as those for the CSS given in ref.\ \cite{2dss},
except with modified Bessel functions $I_\nu $ replaced
by modified spherical Bessel functions $i_\nu $
with different arguments.
However,
wave packets of this type are also unacceptable
because certain physical expectation values diverge.
For example,
$\vev{{\vec L}^2 - L_3^2}$ diverges.

Obtaining an acceptable minimum-uncertainty wave packet
evidently requires setting $\be \ne 0$
and deriving a suitable nonconstant $f(\th)$.
Further physical input is needed to accomplish this.
Note that,
whatever its form,
the function $f(\th)$ must satisfy the limit
$f(\th) \rightarrow 0$ as $\th \rightarrow \pi/2$,
so that $\ch(\th,\ph)$ reduces correctly to a CSS.
Furthermore,
the reflection symmetry about the $x$-$y$ plane
suggests that $f = f(\sin\th)$,
independent of $\cos\th$.

The function $f(\th)$ controls the spread
transverse to the orbit of the initial wave packet.
This spread should depend on the value of $\be =\vev{L_3}$,
since physically as $\vev{L_3}$ increases
the wave packet is expected to become more confined
to a region around the $x$-$y$ plane.
For simplicity,
we take the spread as independent
of the spread around the orbit,
which is controlled by the parameter $\de$.
It turns out that this assumption suffices for our purposes.
If it is relaxed,
the resulting wave packets have a
significantly more complicated analytical structure.
In what follows,
we therefore permit $f(\th)$ to depend on $\be$,
but we require it to be independent of $\de$.

A wave packet following a keplerian orbit
with maximal confinement to the $x$-$y$ plane
should have minimal expectation value
$|\vev{{\vec L}^2 - L_3^2}|$.
We can use this physical condition
to obtain $f(\th)$,
as follows.
Since $f(\th)$ is independent of $\de$,
the limit $\de \rightarrow 0$
can be imposed in Eq.\ \rf{gsss} without loss of generality.
In this limit,
the uncertainty $\De L_3 $ vanishes
because $\ch(\th,\ph)$ becomes an eigenstate
of $L_3$ with eigenvalue $m=\be$.
Expanding $\ch(\th,\ph)$ in terms of spherical harmonics
$Y_{l m} (\th,\ph)$ with $m=\be$ gives
\beq
\ch(\th,\ph) \Bigg\vert_{\de \rightarrow 0}
= \sum_{l \ge \be} c_l \, Y_{l \be} (\th,\ph)
\quad .
\label{yexpans}
\eeq
Then,
calculating the expectation value
$|\vev{{\vec L}^2 - L_3^2}|$
in the state
$\ch(\th,\ph) \Big\vert_{\de \rightarrow 0}$
shows that it is minimized
when the coefficients $c_l$ in Eq.\ \rf{yexpans}
are proportional to $\de_{l \be}$.
The sum reduces to a single term proportional to
$Y_{\be \be}$ and hence proportional to
$\sin^\be \th \exp(i \be \ph)$.
This implies that $\exp(f(\th)) = \sin^\be \th$.
As expected,
$f(\th)$ is a function of $\sin\th$
and satisfies the condition $f(\pi/2) = 0$.
Moreover,
this particular functional form
leads to an SSS without singularities at the poles
and hence with finite physical expectation values.

Combining this result with Eq.\ \rf{gsss}
produces a set of minimum-uncertainty states
satisfying the requirements for SSS.
We therefore take the SSS to be given by
\beq
\ch(\th,\ph) = N \sin^\be \th \exp \left( \de \sin\th \cos\ph
+ i \be \ph \right)
\quad .
\label{sss}
\eeq

\vglue 0.6cm
{\bf\noindent C. Features of SSS}
\vglue 0.4cm

We next calculate the normalization constant
and some expectation values for the SSS.
These are most conveniently specified in terms of a function
$A_j^\be (\de)$,
defined as
\beq
A_j^\be (\de) = 2 \pi \int_0^\pi d\th \, (\sin\th)^{2 \be + j + 1}
\, I_j (2 \de \sin\th)
\quad ,
\label{alanint}
\eeq
where $j$ is an integer
and $I_j(z)$ is a modified Bessel
function of the first kind.
The Appendix presents analytical expressions
for the functions $A_j^\be (\de)$
and discusses some useful properties.

The normalization constant $N$ is given by
\beq
N = \fr 1 {\sqrt{A_0^\be (\de)}}
\quad .
\label{bign}
\eeq
The nonvanishing expectation values of the
angular-coordinate operators are
\beq
\vev{a_1} = \fr {A_1^\be (\de)} {A_0^\be (\de)}
\quad ,
\label{e1}
\eeq
\beq
\vev{a_1^2} = \fr 1 {2 A_0^\be (\de)}
\left( A_0^{\be+1} (\de) + A_2^\be (\de) \right)
\quad ,
\label{e2}
\eeq
\beq
\vev{a_2^2} = \fr 1 {2 \de}
\fr {A_1^\be (\de)} {A_0^\be (\de)}
= \fr 1 {2 \de} \vev{\sin\th \cos\ph}
\quad ,
\label{e3}
\eeq
\beq
\vev{a_3^2} = \fr 1 {A_0^\be (\de)}
\left( A_0^{\be} (\de) - A_0^{\be+1} (\de) \right)
\quad ,
\label{e4}
\eeq
while some useful expectations involving the angular momentum are
\beq
\vev{L_3^2} = \fr \de 2 \fr {A_1^\be (\de)} {A_0^\be (\de)}
+ \be^2
= \fr \de 2 \vev{\sin\th \cos\ph} + \be^2
\quad ,
\label{e5}
\eeq
\bea
\vev{{\vec L}^2} &=& \be(\be+1) - \de^2 \left( 1 -
\vev{(\sin\th \cos\ph)^2} \right)
\nonumber \\
&& \quad\quad\quad + \, \fr {2\de} {A_0^\be (\de)}
\left[ (\be+1) A_1^\be (\de) - \be A_1^{\be-1} (\de) \right]
\quad .
\label{e6}
\eea
Using the identity
\rf{identity}
for the $A_j^\be (\de)$ in the Appendix,
one can show that
$\sum_j a_j^2 = 1$
as required.
Note that,
in the limit $\de \rightarrow 0$,
$\vev{L_3^2} = \be^2$ and $\vev{{\vec L}^2} = \be(\be+1)$,
as is appropriate for an eigenstate with $l=m=\be$.

We next turn to a discussion of some properties of the SSS.
First,
expanding $\ch(\th,\ph)$ in spherical harmonics
for the case with $\de \ne 0$
shows that it is a superposition of both $l$ and $m$ states.
The condition $\exp(f(\th)) = \sin^\be \th$
ensures that the squared modulus $\vert c_{l m} \vert^2$
of the weighting coefficients $c_{l m}$ in the expansion
is maximal for $l=m=\be$ and falls rapidly away from these values.
The rapid decrease indicates that
$\vev{{\vec L}^2 - L_3^2}$ remains small for $\de \ne 0$,
as expected.
The reader interested in the form of the expansion can find
it in Eq.\ \rf{mess} of Sec.\ III.

We have seen that the parameter $\be$ is the expectation
of the $z$ component of the angular momentum.
It turns out that
the parameter $\de$ has an interpretation similar to that
of its counterpart for the CSS.
The uncertainty $\De L_3$ depends on $\de$.
This may be confirmed in the neighborhood of
$x=1$, $y=z=0$ on the unit sphere by expanding the
wave packet around $\ph = 0$ at $\th = \pi/2$.
We find $\vert \chi(\pi/2,\ph) \vert^2 \propto
\exp[ 2 \de (1 - \half \ph^2 + \cdots )]$,
which to leading order is a gaussian in $\ph$
with standard deviation $\si = 1/\sqrt{2 \de}$.
Therefore,
for larger values of $\de$
the initial wave packet becomes narrower in $\ph$.

Classically,
a particle moving along a trajectory
confined to the $x$-$y$ plane has angular motion
depending only on $\ph$.
The trajectory is initialized by giving values
of $\sin\ph$, the sign of $\cos\ph$, and $L_3$.
At the quantum level,
the initial angular position of an SSS
is specified by the expectation values
in Eqs.\ \rf{cond}.
As we have seen,
the parameter $\be$ gives the initial angular momentum
$\vev{L_3}$
and controls the transverse angular spread,
while $\de$ controls the angular spread along the orbit.
Compared to the corresponding classical problem,
a general quantum packet would need two additional parameters
determining the spread on the surface of the sphere.
However,
our simplifying assumption for $f(\th)$ means that
only one extra parameter is needed
to specify the quantum solution for the SSS.
A related point is that
the SSS and CSS
depend on the same number of parameters.
This is physically reasonable since both states follow the
same keplerian orbit in the two-dimensional $x$-$y$ plane.

\vglue 0.6cm
{\bf\noindent III. KEPLERIAN SQUEEZED STATES}
\vglue 0.4cm

This section discusses the keplerian squeezed states.
Their definition is presented in Sec.\ IIIA,
where their parameters are determined in terms
of specified physical quantities.
The time evolution of a KSS is examined in Sec.\ IIIB,
and an example of a KSS wave packet moving along an
elliptical orbit is provided.

\vglue 0.6cm
{\bf\noindent A. Construction and Specification of KSS}
\vglue 0.4cm

The SSS solution
\rf{sss}
is a function only of the angular coordinates
and has neither time dependence nor dependence on the
quantum number $n$.
A solution consisting of a product of an SSS with a
radial energy eigenstate of given $n$ might be considered,
but this produces a stationary state.
However,
we can create a three-dimensional state moving on a
keplerian orbit by combining an SSS with an RSS.

The RSS are constructed and analyzed in
refs.\ \cite{rss,rssqdt}.
They are wave packets localized in the radial coordinates
that initially undergo oscillatory motion between the inner and
outer apsidal points of the corresponding keplerian ellipse.
The period $T_{\rm cl}$ of the motion is that
of a classical particle moving in a Coulomb potential.
The construction involves converting the
classical effective radial hamiltonian for the Coulomb potential
in terms of conventional radial variables
to an oscillator description in terms of new variables.
The resulting classical problem is quantized,
and wave functions are obtained that minimize the
ensuing quantum uncertainty relation.
The RSS are given by
\beq
\psi (r) = N^\prime r^{\al} \exp [-\ga_0 r -i \ga_1 r]
\quad ,
\qquad
N^\prime = [(2 \ga_0 )^{2\al + 3}/\Ga (2 \al + 3)]^{1/2}
\quad .
\label{rss}
\eeq
For the initial wave packet,
the parameters $\al$, $\ga_0$, and $\ga_1$ determine
the radial position,
the uncertainty in the radial variables,
and the radial momentum.
In refs.\ \cite{rss,rssqdt},
the RSS are shown to describe a Rydberg wave packet
that has been excited by a single short laser pulse.
The angular part of the full three-dimensional wave function
in this case is fixed to a p state
for excitation from the ground state.

A minimum-uncertainty wave packet moving on a keplerian
orbit can be obtained by combining an RSS with an SSS.
The resulting wave packet is a KSS.
The separability of the full hamiltonian
and the independence of $l$
of the RSS uncertainty in the new radial coordinate
together make it possible to minimize
simultaneously the uncertainty relations
for the new radial variables
and those for the angular variables in \rf{ur1}.
We therefore can take as an initial wave function the
product of $\psi(r)$ in
\rf{rss}
and $\ch(\th,\ph)$ in
\rf{sss}.
The result is a normalized five-parameter family of KSS,
\bea
\Psi (r,\th,\ph) &=& \psi(r)\chi(\th,\ph)
\nonumber \\
&=&
N N^\prime r^\al \sin^\be \th \,
\exp [{\de \sin\th \cos\ph -(\ga_0 + i \ga_1) r +  i \be \ph}]
\quad ,
\label{ess}
\eea
where $N$ is given by Eq.\ \rf{bign}
and $N^\prime$ is given by Eq.\ \rf{rss}.
The KSS are minimum-uncertainty wave packets
localized in all three dimensions.
The choice of the initial angular-coordinate location is
implicit in the SSS construction and is specified in Eq.\ \rf{cond}.

Expectation values of operators for the KSS
can be calculated analytically.
Since the radial and angular wave functions separate,
the angular operators have the expectation values
given in Eqs.\ \rf{e1} -- \rf{e6}.
The expectation values for the radial operators are
\beq
\vev{r} = \fr {2 \al + 3} {2 \ga_0} ~~,~~~~
\vev{\fr 1 r} = \fr {\ga_0} {\al + 1}
\quad ,
\label{e7}
\eeq
\beq
\vev{r^2} = \fr {(\al + 2)(2 \al + 3)} {2 {\ga_0}^2} ~~,~~~~
\vev{\fr 1 {r^2}} = \fr {2 {\ga_0}^2} {(\al + 1)(2 \al + 1)}
\quad ,
\label{e8}
\eeq
\beq
\vev{p_r} = - \ga_1~~,~~~~
\vev{{p_r}^2} = \fr {{\ga_0}^2} {2 \al + 1} + {\ga_1}^2
\quad .
\label{e9}
\eeq
The uncertainty for the radial coordinates is
\beq
\De r \De p_r = \half \sqrt{\fr {2 \al + 3} { 2 \al + 1}}
\quad .
\label{rp}
\eeq
The RSS are not minimum-uncertainty states in the variables
$r$ and $p_r$,
which is as expected since the construction minimizes the
uncertainty in the new radial variables instead.
For large values of $\al$,
however,
$\De r \De p_r \rightarrow 1/2$.

The expectation value for the energy $\vev{H}$ is obtained
using the full Coulomb hamiltonian and depends on all five
of the KSS parameters.
It is
\beq
\vev{H} = \fr 1 2 \vev{p_r^2} + \fr 1 2 \vev{\fr 1 {r^2}}
\vev{{\vec L}^2} - \vev{\fr 1 r}
\quad .
\label{e10}
\eeq

We next address the issue of initializing a KSS.
With the wave packet located at the outer apsidal point
of an elliptical orbit,
the uncertainty product for a radial Rydberg wave packet
is a minimum.
Initially imposing the constraints \rf{cond}
ensures that the ellipse has semimajor axis aligned along
the $x$ axis.
The five KSS parameters $\al$, $\be$, $\ga_0$, $\ga_1$, and $\de$
can then be fixed by specifying
the expectations of the radial coordinates
$\vev{r}$ and $\vev{p_r}$,
the expectation of the angular momentum $\vev{L_3}$,
the expectation of the energy $\vev{H}$,
and the spread $\De L_3$ in the angular momentum.

The natural choice for
the initial expectation value of $p_r$ is zero.
Similarly,
the natural choice for the initial expectation value of $r$
is the outer apsidal point of the orbit,
$r_{\rm out} = n^2 \left(1 + \sqrt{(1 - l(l+1)/n^2} \right)$.
The expectation $\vev{L_3}$ gives the initial angular momentum
of the wave packet and its spread transverse to the orbit.
The two remaining conditions fix the initial spread
of the wave packet in the radial and orbital directions.
For the first,
a natural choice is to set $\vev{H}$ equal to the
mean energy of a Rydberg wave packet consisting of a
superposition of $n$ states centered on the value $\bar n$.
A packet of this type is produced by excitation with a short
laser pulse tuned to the mean energy
$E_{\bar n} = {-1}/{2 {\bar n}^2}$.
For the second,
it suffices to specify the width $\De L_3$ of the superposition.
Note, however, that a single laser pulse
cannot excite a wave packet localized in the angular coordinates.
Creation of a superposition of angular eigenstates requires
additional fields to mix angular-momentum eigenstates.

The full set of conditions sufficient to fix the five KSS parameters
are then
\bea
\vev{r} = r_{\rm out}
\quad , \qquad
\vev{p_r}  &=& 0
\quad , \qquad
\vev{L_3} = \be
\quad ,
\nonumber \\
\vev{H} = E_{\bar n}
\quad , \qquad
&&
\sqrt{\vev{L_3^2} - \vev{L_3}^2} = \De L_3
\quad .
\label{params}
\eea
These determine the values of $\al$, $\be$, $\ga_0$,
$\ga_1$, and $\de$ in terms of physical parameters
in the excitation process.

\vglue 0.6cm
{\bf\noindent B. Evolution of KSS}
\vglue 0.4cm

We next investigate the time evolution of
a KSS wave function matched to a Rydberg wave
packet at the outer apsidal point of an elliptical orbit,
as described above.
By construction,
the packet is expected to travel along
a classical ellipse in the $x$-$y$ plane.

Since $\vev{p_r} = 0$ and $\vev{L_3} = \be$,
the initial motion of the wave packet
for $\be > 0$ is in the direction of
increasing $\ph$.
The geometry of the ensuing orbit depends on the values of
$\bar n$, $\be$, and $\De L_3$.
The parameter $\be$ gives the central values of $l$ and $m$.
For $\be \simeq {\bar n} - 1$,
the KSS orbit becomes circular,
with the wave packet propagating at fixed mean radial
distance from the origin.
As $\be$ decreases,
the orbit becomes elliptical,
with the inner apsidal point moving closer to the nucleus.
Also,
for $\be \rightarrow 1$
the radial wave function becomes oscillatory as the
electron passes close to the nucleus.

To study the time evolution of a KSS,
we expand $\Psi (r,\th,\ph)$ in
\rf{ess}
in terms of energy and angular-momentum eigenstates,
\beq
\Psi(r,\th,\ph,t) = \sum_{n,l,m} c_{nlm} R_{nl}(r) Y_{lm} (\th,\ph)
e^{-i E_n t}
\quad .
\label{expans}
\eeq
The expansion coefficients
$c_{nlm} = \vev{\Psi(r,\th,\ph,0) \vert R_{nl}(r) Y_{lm} (\th,\ph)}$
can be calculated using
Eq.\ \rf{ess}
as the initial wave function $\Psi(r,\th,\ph,0)$.
The radial and angular parts separate,
and we may write $c_{nlm} = c_{nl}^{\rm (rss)} c_{lm}^{\rm (sss)}$,
where $c_{nl}^{\rm (rss)} = \vev{\ps(r) \vert R_{nl}(r)}$ for
$\ps(r)$ in
\rf{rss}
and $c_{lm}^{\rm (sss)} = \vev{\ch(\th,\ph) \vert Y_{lm} (\th,\ph)}$
for $\ch(\th,\ph)$ in
\rf{sss}.
The coefficients $c_{nl}^{\rm (rss)}$ for the radial part of
the expansion are given in Eq.\ (58) of
ref.\ \cite{rssqdt}.
The coefficients for the angular expansion can be calculated using
\rf{sss}.
The result for $m \ge 0$ is
\bea
c_{lm}^{\rm (sss)} &=& N 4 \sqrt{\pi} \,
\de_{0,(l-m)_{\rm mod \, 2}} \, (-1)^m
\sqrt{\fr {2l+1} {4 \pi} \fr {(l-m)!} {(l+m)!}}
\, \, \sum_{k=0}^{\fr 1 2 (l-m)} \sum_{p=0}^{{\rm min}(m,\be)} \,
\fr {(-1)^k} {2^k k!}
\nonumber \\
&&
\times \, \fr {(2l-2k-1)!!} {(l-m-2k)!} (-1)^p
\left(\matrix{{\rm min}(m,\be)\cr p\cr}\right) \,
2^{\fr 1 2 (l-m+2p-2k)}
\nonumber \\
&&
\quad\quad
\times \,
\fr {\Ga(\fr 1 2 (l-m+2p-2k+1))}
{\de^{\fr 1 2 (l-m+2p-2k)}} \, \,
i_{\fr 1 2 (2\be+l-3m+2p-2k)}(\de)
\quad .
\label{mess}
\eea
Here,
$N$ is the SSS normalization constant in \rf{bign},
${\rm min}(m,\be)$ gives the minimum of the two values $m$ and $\be$,
and $i_\nu (z)$ is a modified spherical Bessel function.
The appearance of the Kronecker delta
$\de_{0,(l-m)_{\rm mod \, 2}}$
implies that if $l-m$ is an odd integer then $c_{lm}^{\rm (sss)} = 0$.
Therefore,
the SSS is composed only of eigenstates for which $l-m$ is even.
This follows because the associated Legendre functions
with odd $l-m$ are odd functions of $\th$
in the interval $0 \le \th \le \pi$ centered on $\pi/2$,
while $\ch(\th,\ph)$ is an even function of $\th$
in this interval.

As an explicit example of the evolution of a KSS,
consider matching $\Psi(r,\th,\ph,0)$
to a Rydberg wave packet at the outer apsidal
point with $\bar n = 45$,
$\vev{L_3} = 30$,
and $\De L_3 = 2.5$.
Using these values and
Eqs.\ \rf{params},
we obtain the KSS-parameter values $\al \simeq 62.846$,
$\be = 30$, $\ga_0 \simeq 0.01834$, $\ga_1 = 0$, and
$\de \simeq 12.826$.
This gives $\vev{r} = r_{\rm out} \simeq 3508.6$ a.u.\
and $\vev{{\vec L}^2} \simeq 938.1$.
Defining a mean value $\bar l$ of $l$
from the relation $\vev{{\vec L}^2} = {\bar l}({\bar l}+1)$,
we find $\bar l \simeq 30.1 \approx \be$.

The series in
\rf{expans}
may be well approximated by truncating the sum to a finite
number of terms with $n$ centered on $\bar n$ and $l$ and $m$
centered on $\be$.
In the present example,
we allow a spread of 10 units in $n$ and $l$ and four
units in $m$.
This gives 484 coefficients, half of which vanish.
We keep the remaining 242 terms in the series and plot two-dimensional
sections through the KSS as a function of $t$.

Figure 1 presents a slice through the initial KSS in
the $x$-$z$ plane.
This slice is transverse to the classical orbit,
which lies in the $x$-$y$ plane.
The figure shows that the initial wave packet is localized
around a point on the $x$ axis at $r \approx r_{\rm out}$
and in a narrow range of $\th$ near $\th = \pi/2$,
corresponding to $z \approx 0$.

Figure 2a presents a slice through the initial wave packet
in the plane of the classical orbit.
The initial wave packet is located on the positive $x$ axis
at the outer apsidal point.
Figures 1 and 2a taken together show that the initial KSS is
localized in all three dimensions.

The classical keplerian orbit for a particle in a Coulomb potential
is $T_{\rm cl} = 2 \pi {\bar n}^3$.
With $\bar n = 45$,
we obtain $T_{\rm cl} \simeq 13.4$ psec.
Figure 2b shows the KSS in the $x$-$y$ plane at $t =  T_{\rm cl}/3$.
It has moved in the direction of positive $\ph$ and is spreading
along the elliptical orbit.
In accordance with Kepler's laws,
the wave packet moves more slowly near the outer apsidal point
than near the inner one.
As a result,
it has traveled less than $1/3$ of the orbital circumference
at $t = T_{\rm cl}/3$.

Figure 2c shows the KSS in the $x$-$y$ plane at $t =  T_{\rm cl}/2$.
It has spread along the elliptical orbit and is moving more rapidly.
The radial distance to the inner apsidal point is
$r_{\rm in} \simeq 536$ a.u.
This is sufficiently far from the nucleus for the wave packet
to remain localized in $r$,
and hence no radial oscillations are apparent.

Figures 2d and 2e show the slice through
the KSS in the $x$-$y$ plane
at the times $t =2T_{\rm cl}/3$ and $t  = T_{\rm cl}$,
respectively.
The motion slows again as the wave packet approaches
the outer apsidal point
and becomes more localized.
At $t = T_{\rm cl}$,
the wave packet resembles the initial wave packet.
However,
the motion is not exactly periodic.
As time increases,
the wave packet collapses and for $t \gg T_{\rm cl}$ a
cycle of revivals and superrevivals commences.

Figure 3 shows the wave packet in the $x$-$y$ plane
at $t = T_{\rm cl}/2$,
but viewed from a point on the positive $x$ axis looking towards
the nucleus.
The elliptical shape of the orbit is evident.

\vglue 0.6cm
{\bf\noindent IV. INCORPORATION OF QUANTUM DEFECTS}
\vglue 0.4cm

Experiments studying the behavior of Rydberg wave packets are
usually performed using alkali-metal atoms.
These have energies given by the Rydberg series
$E_{n^\ast} = {-1}/{2n^\ast}$,
where $n^\ast = n - \de(n,l)$,
and $\de(n,l)$ is a quantum defect.
The empirical parameters $\de(n,l)$ give the
energy-level shifts away from hydrogenic values.
For large $n$,
they approach asymptotic values $\de(l)$ independent of $n$.

In ref.\ \cite{rssqdt},
it is shown that the RSS construction can be generalized
to include the effects of quantum defects.
The analysis uses a model called supersymmetry-based
quantum-defect theory (SQDT) to describe alkali-metal atoms
\cite{sqdt}.
This analytical theory for alkali-metal atoms
has exact asymptotic quantum-defect energies as eigenvalues.
The SQDT eigenfunctions form a complete and orthogonal set.

The SQDT eigenstates are $R_{n^\ast l^\ast}(r) Y_{lm}(\th,\ph)$,
where $n^\ast = n - \de(l)$, $l^\ast = l - \de(l) + I(l)$,
and $I(l)$ is an integer that depends on $l$.
The radial eigenstates $R_{n^\ast l^\ast}(r)$ have the same
functional form as the hydrogenic functions $R_{n l}(r)$,
but $n$ is replaced by $n^\ast$ and $l$ is replaced by $l^\ast$.
The angular wave functions are the usual eigenstates
$Y_{lm}(\th,\ph)$ of the angular momentum.

Our KSS construction in three dimensions can be
generalized to include the effects of quantum defects
\cite{fn}.
Since the angular part of the solution in SQDT separates,
the SSS wave functions remain unchanged and are given by
$\ch(\th,\ph)$ in
Eq.\ \rf{sss}.
The RSS wave functions $\ps(r)$ are obtained by
writing the classical SQDT hamiltonian in terms of
new radial oscillator variables incorporating
effects of quantum defects
and then finding minimum-uncertainty
solutions for the corresponding quantum problem.
The resulting wave function $\ps(r)$ has a related
functional form to that in Eq.\ \rf{rss}
and is discussed in ref.\ \cite{rssqdt}.

A KSS $\Psi(r,\th,\ph)$ for alkali-metal atoms
is again formed as a product of an
RSS $\ps(r)$ and an SSS $\ch(\th,\ph)$.
The functional form of the solution is similar to that in
Eq.\ \rf{ess}
but with suitable replacements for the quantum numbers.
To allow for the shifted energy eigenvalues
of the alkali-metal atoms,
the initialization procedure for the parameters must be modified.
We choose $\vev{p_r} = 0$ and $\vev{L_3} = \be$,
and specify $\De L_3$ as before.
Denote by $E_{{\bar n}^\ast}$ the energy expectation
of the wave packet in the excited alkali-metal atom,
and let $r_{\rm out}^\ast$ be the outer apsidal point for a
superposition of states with quantum-defect eigenenergies.
Then,
we impose
$\vev{H} = E_{{\bar n}^\ast}$ and $\vev{r} = r_{\rm out}^\ast$,
which differ from the hydrogenic case.
To calculate $\vev{H}$ explicitly,
we can take advantage of the completeness of the SQDT eigenfunctions
and expand the initial KSS as a superposition of SQDT eigenstates,
\beq
\Psi(r,\th,\ph,0) = \sum_{n,l,m} {\tilde c}_{nlm}
R_{n^\ast l^\ast}(r) Y_{lm} (\th,\ph)
\quad ,
\label{sqdtexpans}
\eeq
where the expansion coefficients,
which depend on the KSS parameters,
can be determined by inversion.
The expectation value for the hamiltonian is then
specified by
\beq
\vev{H} = \sum_{n,l,m} \vert {\tilde c}_{nlm} \vert^2 E_{n^\ast}
= E_{{\bar n}^\ast}
\quad .
\label{hexp}
\eeq

It is known that the long-term revival times for an
alkali-metal wave packet depend on the quantum defects and
that the effects of the quantum defects are different from
the effects of a laser detuning
\cite{detunings}.
In addition,
the appearance of deviations from the hydrogenic potential
arising from the presence of core electrons in an alkali-metal atom
means that the Runge-Lenz operator $\vec A$ does not commute
with the hamiltonian.
The classical orbit therefore precesses at a rate determined by the
quantum defect
\cite{rssqdt}.

\vglue 0.6cm
{\bf\noindent V. SUMMARY}
\vglue 0.4cm

In this paper,
we have obtained minimum-uncertainty wave-packet solutions
for the Schr\"odinger equation with a Coulomb potential in
three dimensions.
The solutions are the KSS and are given as a product of RSS with SSS.
The RSS,
previously derived,
minimize the uncertainty relation for radial variables
expressing the radial Coulomb problem in oscillator form.
The SSS,
constructed here,
minimize the uncertainty relation for angular-coordinate
and angular-momentum operators.
The KSS provide analytical solutions to the Coulomb problem
that move along classical keplerian orbits.
They exhibit both classical and quantum-mechanical features.

The KSS can be used as an analytical tool for studying the
quantum-classical correspondence in the Coulomb problem.
They may also be used to describe Rydberg wave packets created
by excitation of a Rydberg atom with a short laser pulse in the
presence of external fields.
Such wave packets are expected to move in three dimensions
along elliptical orbits that are strongly peaked around a plane.
To match a KSS to a Rydberg wave packet,
we choose the outer apsidal point as the initial location
of the wave packet.
The five KSS parameters are determined from the expectation
values of the radial position,
the radial momentum,
the energy,
the angular momentum $L_3$ transverse to the orbital plane,
and the spread in $L_3$.

We obtained the time evolution of a KSS
and provided an explicit example.
The wave packet moves along an elliptical orbit with the classical
keplerian orbital period.
The width of the wave packet oscillates during the motion,
as is characteristic of a squeezed state.
The KSS maintain their shape for several orbital cycles
before collapsing and undergoing quantum recurrences.

Finally,
we provided an extension of the construction to the
case where quantum defects are present.
With this analysis,
the KSS can be used for the description
of wave packets in alkali-metal atoms,
which are the ones of choice in current experiments.

\vglue 0.6cm
{\bf\noindent ACKNOWLEDGMENTS}
\vglue 0.4cm

This work is supported in part by the National
Science Foundation under grant number PHY-9503756.

\vglue 0.6cm
{\bf\noindent APPENDIX}
\vglue 0.4cm

In this appendix,
we discuss the functions $A_j^\be (\de)$ defined in
Eq.\ \rf{alanint}
and examine some of their properties.

The integral in Eq.\ \rf{alanint}
can be evaluated.
We find
\beq
A_j^\be (\de) = 4 \sqrt{\pi} \sum_{k=0}^\be
\fr {(-1)^k} {\de^k}
\left(\matrix{\be\cr k\cr}\right)
\Ga (k + \fr 1 2) \, i_{j+k} (2 \de)
\quad ,
\label{alan}
\eeq
where $i_n (z)$ is a modified spherical Bessel function.
This exact expression permits the numerical computation
of $A_j^\be (\de)$ to arbitrary precision using standard procedures.

A useful identity for the $A_j^\be (\de)$ can be derived
by taking advantage of some properties
of the modified spherical Bessel functions.
We obtain
\beq
A_j^{\be + 1} (\de) = A_{j+2}^\be (\de)
+ \fr {j+1} \de A_{j+1}^\be (\de)
\quad .
\label{identity}
\eeq
For $j>0$,
we find $A_j^\be (0) = 0$.
For $j=0$,
however,
the value of the function with zero argument is
\beq
A_0^{\be} (0) = \fr {4 \pi (2 \be)!!} {(2 \be +1)!!}
\quad .
\label{a0}
\eeq
These relations are used to simplify some expressions in
the main body of the text.

\vfill
\newpage

{\bf\noindent REFERENCES}
\vglue 0.4cm

\vfill\eject

\baselineskip=16pt
{\bf\noindent FIGURE CAPTIONS}
\vglue 0.4cm

\begin{description}

\item[{\rm Fig.\ 1:}]
A slice through the initial KSS wave packet
in the $x$-$z$ plane,
which is transverse to the plane of the classical orbit.
The quantity $r^2 \vert \Psi(r,\th,\ph) \vert^2$ (in arbitrary units)
is plotted as a function of $x$ and $z$ at $t=0$.

\item[{\rm Fig.\ 2:}]
Slices through the KSS wave packet in the $x$-$y$ plane,
at different times during the classical orbital cycle.
The quantity
$r^2 \vert \Psi(r,\th,\ph,t) \vert^2$ (in arbitrary units)
is shown as a function of $x$ and $y$ at the times
(a) $t=0$,
(b) $t = \fr 1 3 T_{\rm cl}$,
(c) $t = \fr 1 2 T_{\rm cl}$,
(d) $t = \fr 2 3 T_{\rm cl}$,
(e) $t = T_{\rm cl}$.

\item[{\rm Fig.\ 3:}]
The KSS wave packet at time $t = \fr 1 2 T_{\rm cl}$
shown in Fig.\ 2c,
viewed from a point on the positive $x$ axis
looking toward the nucleus.

\end{description}

\end{document}